\documentclass[table]{article}

\usepackage[utf8]{inputenc} 
\usepackage[T1]{fontenc}    
\usepackage{xcolor}
\definecolor{TUMblue}{HTML}{3070B3}

\usepackage[colorlinks=true,allcolors=TUMblue]{hyperref}
\usepackage[capitalize,noabbrev]{cleveref}
\usepackage[
    backend=biber,
    style=authoryear-comp,
    maxcitenames=1,
    uniquename=false,
    sortcites=ynt,
    uniquelist=false,
    natbib=true,
    hyperref=true,
    maxbibnames=10,
    giveninits=true,
    uniquename=init,
    url=false,
    isbn=false,
    doi=false
 ]{biblatex}
\usepackage{acro}
\acsetup{list/template=description}
\DeclareAcronym{RMSD}{
	short = RMSD,
	long = root mean square deviation of atomic positions,
}
\DeclareAcronym{qHTS}{
	short = qHTS,
	long = quantitative high-throughput screening,
}

\DeclareAcronym{LLM}{
	short = LLM,
	long = large language model,
}

\DeclareAcronym{PLM}{
	short = PLM,
	long = protein language model,
}

\DeclareAcronym{MT-DNN}{
	short = MT-DNN,
	long = multi-task deep neural network,
}

\DeclareAcronym{jpg}{
	short = JPEG ,
	sort = jpeg ,
	alt = JPG ,
	long = Joint Photographic Experts Group
}

\DeclareAcronym{ML}{
	short = ML,
	long = machine learning
}

\DeclareAcronym{DL}{
	short = DL,
	long = deep learning
}

\DeclareAcronym{MCC}{
	short = MCC,
	long = Matthews correlation coefficient
}

\DeclareAcronym{Sp}{
	short = Sp,
	long = specificity
}

\DeclareAcronym{Sn}{
	short = Sn,
	long = sensitivity
}

\DeclareAcronym{BA}{
	short = BA,
	long = balanced accuracy
}

\DeclareAcronym{AP}{
	short = AP,
	long = average precision
}

\DeclareAcronym{BEDROC}{
	short = BEDROC,
	long = Boltzmann-enhanced discrimination of receiver operating characteristic
}

\DeclareAcronym{ROC_AUC}{
	short = ROC AUC,
	long = receiver operating characteristic area under curve
}

\DeclareAcronym{PR_AUC}{
	short = PR AUC,
	long = precision recall area under curve
}

\DeclareAcronym{DPR_AUC}{
	short = $\Delta$PR AUC,
	long = $\Delta$ in precision recall area under curve
}

\DeclareAcronym{TPR}{
	short = TPR,
	long = true positive rate
}

\DeclareAcronym{TNR}{
	short = TNR,
	long = true negative rate
}

\DeclareAcronym{FPR}{
	short = FPR,
	long = false positive rate
}

\DeclareAcronym{FNR}{
	short = FNR,
	long = false negative rate
}

\DeclareAcronym{TP}{
  short = TP,
  long  = true positive
}

\DeclareAcronym{FN}{
  short = FN,
  long  = false negative
}

\DeclareAcronym{FP}{
  short = FP,
  long  = false positive
}

\DeclareAcronym{TN}{
  short = TN,
  long  = true negative
}

\DeclareAcronym{RF}{
	short = RF,
	long = random forest
}

\DeclareAcronym{AID}{
	short = AID,
	long = bioassay identifier
}

\DeclareAcronym{HTS}{
	short = HTS,
	long = high-throughput screening
}

\DeclareAcronym{MMP}{
	short = MMP,
	alt = $\Delta\Psi_\text{m}$,
	long = mitochondrial membrane potential 
}

\DeclareAcronym{m-MPI}{
	short = m-MPI,
	long = mitochondrial membrane potential indicator
}

\DeclareAcronym{ECACC}{
	short = ECACC,
	long = European Collection of Authenticated Cell Cultures 
}

\DeclareAcronym{DMEM}{
	short = DMEM,
	long = Dulbecco's modified eagle medium 
}

\DeclareAcronym{FCS}{
	short = FCS,
	long = fetal calf serum
}

\DeclareAcronym{RT}{
	short = RT,
	long = room temperature
}

\DeclareAcronym{FCCP}{
	short = FCCP,
	long = carbonylcyanid-4-(trifluormethoxy)phenylhydrazon,
}

\DeclareAcronym{DMSO}{
	short = DMSO,
	long = dimethyl sulfoxide,
}

\DeclareAcronym{ddH2O}{
	short = \ch{ddH2O},
	long = double destilled water,
	sort={ddH2O},
}

\DeclareAcronym{PBS}{
	short = PBS,
	long = phosphate-buffered saline,
}

\DeclareAcronym{EC50}{
	short = EC\textsubscript{50},
	long = half maximal effective concentration,
}

\DeclareAcronym{AI}{
	short = AI,
	long = artificial intelligence,
}

\DeclareAcronym{DTI}{
	short = DTI,
	long = drug--target interaction,
}

\DeclareAcronym{DDI}{
	short = DDI,
	long = drug--drug interaction,
}

\DeclareAcronym{DNA}{
	short = DNA,
	long = deoxyribonucleic acid,
}

\DeclareAcronym{ECFP}{
	short = ECFP,
	long = extended-connectivity fingerprint,
}

\DeclareAcronym{FCFP}{
	short = FCFP,
	long = functional-class fingerprint,
}

\DeclareAcronym{MAP4}{
	short = MAP4,
	long = MinHashed atom-pair fingerprint,
}

\DeclareAcronym{SVM}{
	short = SVM,
	long = support vector machine,
}

\DeclareAcronym{DNN}{
	short = DNN,
	long = deep neural network,
}

\DeclareAcronym{GCNN}{
	short = GCNN,
	long = graph convolutional neural networks,
}

\DeclareAcronym{GHS}{
	short = GHS,
	long = globally harmonized system of classification and labelling of chemicals,
}

\DeclareAcronym{SMILES}{
	short = SMILES,
	long = simplified molecular-input line-entry system,
}

\DeclareAcronym{CLI}{
	short = CLI,
	long = command-line interpreter,
}

\DeclareAcronym{GUI}{
	short = GUI,
	long = graphical user interface,
}

\DeclareAcronym{ATP}{
	short = ATP,
	long = adenosine triphosphate,
}

\DeclareAcronym{AMP}{
	short = AMP,
	long = adenosine monophosphate,
}

\DeclareAcronym{PPi}{
	short = PP\textsubscript{i},
	long = pyrophosphate,
	sort={PPi}
}

\DeclareAcronym{Lu}{
	short = \ch{LH2},
	long = luciferin,
	sort={LH2}
}

\DeclareAcronym{oLu}{
	short = \ch{oxy-L},
	long = oxy-luciferin,
	sort={oxylu}
}

\DeclareAcronym{SMOTE}{
	short = SMOTE,
	long = synthetic minority over-sampling technique,
}

\DeclareAcronym{SHAP}{
	short = SHAP,
	long = Shapley additive explanation,
}

\DeclareAcronym{CPU}{
	short = CPU,
	long = central processing unit,
}

\DeclareAcronym{RAM}{
	short = RAM,
	long = random-access memory,
}

\DeclareAcronym{GPU}{
	short = GPU,
	long = graphics processing unit,
}

\DeclareAcronym{CAS}{
	short = CAS,
	long = Chemical Abstracts Service,
}

\DeclareAcronym{UMAP}{
	short = UMAP,
	long = uniform manifold approximation and projection,
}

\DeclareAcronym{Tox21}{
	short = Tox21,
	long = Toxicology in the 21\textsuperscript{st} Century,
}

\DeclareAcronym{GOSS}{
	short = GOSS,
	long = gradient-based one-side sampling,
}

\DeclareAcronym{QSAR}{
	short = QSAR,
	long = quantitative structure--activity relationship,
}

\DeclareAcronym{EFB}{
	short = EFB,
	long = exclusive feature bundling,
}

\DeclareAcronym{MTT}{
	short = MTT,
	long = \iupac{3-(4, 5-dimethylthiazolyl-2)-2,5-diphenyltetrazolium bromide},
}

\DeclareAcronym{SSL}{
	short = SSL,
	long = self-supervised learning,
}

\DeclareAcronym{GBM}{
	short = GBM,
	long = gradient boosting machine,
}

\DeclareAcronym{MLP}{
	short = MLP,
	long = multilayer perceptron,
}

\DeclareAcronym{API}{
	short = API,
	long = application programming interface,
}

\DeclareAcronym{CHOP}{
  short = CHOP,
  long  = C/EBP homologous protein
}
\DeclareAcronym{UPR}{
  short = UPR,
  long  = unfolded protein response
}
\DeclareAcronym{ER}{
  short = ER,
  long  = endoplasmic reticulum
}

\DeclareAcronym{PN}{
  short = PN,
  long  = prototypical network
}

\DeclareAcronym{FH}{
  short = FH,
  long  = frequent hitters
}

\DeclareAcronym{LSA}{
  short = LSA,
  long  = latent semantic analysis
}

\DeclareAcronym{ADMET}{
  short = ADMET,
  long  = {absorption, distribution, metabolism, excretion and toxicity}
}

\DeclareAcronym{GNN-ST}{
  short = GNN-ST,
  long  = single-task graph neural network
}

\DeclareAcronym{GNN-MT}{
  short = GNN-MT,
  long  = multi-task graph neural network
}

\DeclareAcronym{GNN-MAML}{
  short = GNN-MAML,
  long  = model-agnostic meta-learning graph neural network
}

\DeclareAcronym{MAT}{
  short = MAT,
  long  = molecule attention transformer
}

\DeclareAcronym{1D}{
    short = 1D,
    long  = one-dimensional 
}

\DeclareAcronym{2D}{
    short = 2D,
    long  = two-dimensional 
}

\DeclareAcronym{3D}{
    short = 3D,
    long  = three-dimensional 
}

\DeclareAcronym{ITC}{
    short = ITC,
    long  = isothermal titration calorimetry
}

\DeclareAcronym{LA}{
    short = LA,
    long  = lipoic acid
}

\DeclareAcronym{CLM}{
    short = CLM,
    long  = chemical language model
}

\DeclareAcronym{RAscore}{
    short = RAscore,
    long  = retrosynthetic accessibility score
}

\DeclareAcronym{AutoML}{
    short = AutoML,
    long  = automated machine learning
}

\DeclareAcronym{SAR}{
    short = SAR,
    long  = structure--activity relationship
}

\DeclareAcronym{FCD}{
    short = FCD,
    long  = Fréchet ChemNet distance
}

\DeclareAcronym{MoA}{
    short = MoA,
    long  = mode of action
}

\DeclareAcronym{ROS}{
    short = ROS,
    long  = reactive oxygen species
}

\DeclareAcronym{VitC}{
    short = VitC,
    long  = vitamin c
}

\DeclareAcronym{GSH}{
    short = GSH,
    long  = glutathione
}

\DeclareAcronym{MIC}{
    short = MIC,
    long  = minimal inhibitory concentration
}

\DeclareAcronym{D2B}{
    short = D2B,
    long  = direct-to-biology
}

\DeclareAcronym{REOS}{
    short = REOS,
    long  = rapid elimination of swill
}

\DeclareAcronym{MCS}{
    short = MCS,
    long  = maximum common substructure
}

\DeclareAcronym{SBDD}{
    short = SBDD,
    long  = structure-based drug design
}

\DeclareAcronym{GNN}{
    short = GNN,
    long  = graph neural network
}

\DeclareAcronym{GPT}{
    short = GPT,
    long  = generative pre-trained transformer
}

\DeclareAcronym{LDDT}{
    short = LDDT,
    long  = local distance difference test
}

\DeclareAcronym{TM}{
    short = TM,
    long  = template modeling
}

\usepackage[preprint,nonatbib]{neurips_2024}

\usepackage{adjustbox}
\usepackage{booktabs}       
\usepackage{amsfonts}       
\usepackage{nicefrac}       
\usepackage{microtype}      
\usepackage{orcidlink}      
\usepackage{listings}
\usepackage[round-mode=places,round-precision=2,group-separator={\,},output-decimal-marker={.},round-pad=false]{siunitx}

\definecolor{maroon}{cmyk}{0,0.87,0.68,0.32}

\newcommand\blfootnote[1]{%
  \begingroup
  \renewcommand\thefootnote{}\footnotetext{#1}%
  \addtocounter{footnote}{-1}%
  \endgroup
}

\title{AI-guided Antibiotic Discovery Pipeline from Target Selection to Compound Identification}

\author{%
  Maximilian G. Schuh\,\orcidlink{0009-0008-2415-8810}$^{1*}$ \quad
  Joshua Hesse\,\orcidlink{0009-0004-1819-7726}$^{1*}$ \quad
  Stephan A. Sieber\,\orcidlink{0000-0002-9400-906X}$^{1}$\\
  $^1$ Chair of Organic Chemistry II\\
  TUM School of Natural Sciences\\
  Technical University of Munich\\
}

\begin{document}

\maketitle

\blfootnote{*Equal contribution.}
\blfootnote{Corresponding author: \href{mailto:stephan.sieber@tum.de}{\texttt{stephan.sieber@tum.de}}}


\begin{abstract}
Antibiotic resistance presents a growing global health crisis, demanding new therapeutic strategies that target novel bacterial mechanisms. 
Recent advances in protein structure prediction and \acl{ML}-driven molecule generation offer a promising opportunity to accelerate drug discovery. 
However, practical guidance on selecting and integrating these models into real-world pipelines remains limited. 
In this study, we develop an end-to-end, \acl{AI}-guided antibiotic discovery pipeline that spans target identification to compound realization. 
We leverage structure-based clustering across predicted proteomes of multiple pathogens to identify conserved, essential, and non-human-homologous targets. 
We then systematically evaluate six leading 3D-structure-aware generative models---spanning diffusion, autoregressive, \acl{GNN}, and language-model architectures---on their usability, chemical validity, and biological relevance. 
Rigorous post-processing filters and commercial analogue searches reduce over \num{100000} generated compounds to a focused, synthesizable set. 
Our results highlight DeepBlock and TamGen as top performers across diverse criteria, while also revealing critical trade-offs between model complexity, usability, and output quality. 
This work provides a comparative benchmark and blueprint for deploying \acl{AI} in early-stage antibiotic development.
\end{abstract}

\acresetall

\section{Introduction}
The integration of \ac{AI} into drug discovery has profoundly reshaped how researchers approach the development of new therapeutics \parencite{zhang_artificial_2025}. \Ac{DL}-driven methodologies offer opportunities to efficiently navigate complex chemical and biological spaces, enabling accelerated discovery and enhanced specificity of therapeutic candidates. Despite these advancements, antibiotic discovery remains particularly challenging, marked by stagnation and limited success in introducing novel antibacterial classes \parencite{farha_important_2025}. Historically, rational drug design---targeting specific bacterial proteins and biological mechanisms---held significant initial promise \parencite{silver_appropriate_2016}. However, limited successes due to restricted uptake of designed compounds shifted attention towards \ac{HTS}, which became the prevailing methodology due to its unbiased, expansive approach \parencite{macarron_impact_2011}. However, \ac{HTS} campaigns are resource-intensive, prone to high false-positive rates, and require extensive follow-up to interpret phenotypic effects---factors that have contributed to their limited success in discovering novel antibiotics \parencite{thorne_apparent_2010, reck2019challenges}.

Recent advances in protein structure prediction and \ac{ML}-driven molecule generation offer compelling reasons to revisit rational drug design \parencite{abramson2024accurate}. Traditional target identification methods, e.g., to identify conserved proteins among pathogenic bacteria, relied heavily on sequence alignment, often missing biologically crucial conserved structural motifs. The emergence of structural alignment methods, notably Foldseek, addresses these limitations by accurately identifying structurally conserved protein regions and efficiently clustering structures on a proteomic scale \parencite{vankempen2024fasta, mifsud2024mapping}.

Despite these technological advancements, practical guidance on the selection, application, and integration of \ac{DL}-driven tools into drug discovery workflows remains limited. To address this gap, we developed a comprehensive and accessible \ac{DL}-guided pipeline designed to facilitate the use of these cutting-edge methods by researchers. Our goal was to provide an unbiased assessment of various \ac{DL}-driven methodologies, outlining clear guidelines on method selection, integration strategies, and practical considerations for researchers seeking to leverage \ac{DL} in their own drug discovery efforts.

In this paper, we develop and validate this comprehensive, \ac{DL}-guided drug discovery pipeline, demonstrating its effectiveness in antibiotic discovery as a proof-of-concept application. The pipeline spans target identification to compound realization, leveraging structure-based clustering of predicted proteomes from multiple pathogens to identify conserved, essential bacterial targets devoid of human homologs \parencite{ji2001identification,salama2004global,chaudhuri2009comprehensive,zhang2012global,turner2015essential,goodall2018essentiala}. 
Additionally, we systematically evaluate six state-of-the-art 3D structure-aware generative models---including diffusion, autoregressive, \ac{GNN}, and language-model architectures---to assess their usability, chemical validity, and biological relevance \parencite{li2024deep, zhang_resgen_2023, schneuing2024structurebased, peng2022pocket2mol, guan20233d, wu_tamgen_2024}. 
Rigorous filtering and commercial analogue searches allow us to refine over \num{100000} generated compounds into a focused, synthesizable subset \parencite{walters_virtual_1998, walters1999recognizing}. 
To facilitate the translation of these computational predictions into tangible compounds, we further screened a vast chemical space containing approximately 5.3 trillion molecules, enabling the efficient selection of candidates that are not only biologically promising but also commercially accessible for cost effective synthesis. 
Our comparative analysis highlights critical trade-offs among model complexity, usability, and output quality, ultimately identifying top-performing models such as DeepBlock and TamGen. 
Through this work, we provide both a benchmark and practical blueprint for integrating \ac{DL} into early-stage drug discovery pipelines.

\subsection{Prior Work}

\subsubsection{Structural Prediction}

AlphFold and RoseTTAFold All-Atom have revolutionized the field of protein structure elucidation \parencite{abramson2024accurate,krishna2024generalized}. 
As a result, the entire proteome of a species can now be analyzed structurally due to predicted protein structures, which would have previously required the extremely laborious experimental elucidation via, e.g., X-ray crystallography. 
This new AlphaFold protein structure database now enables bioinformatic analyses previously only possible on genome level, such as the alignment and clustering of related protein structures across different species \parencite{varadi2022alphafold,varadi2024alphafold}. 
However, structural alignment tools are far slower compared to sequential alignment, both due to the lack of prefilters often used in sequential alignment, and the non-locality of structural similarity scores: Unlike sequence alignments, changes in local structural alignments have effects on the similarity of all other sections of the alignment \parencite{vankempen2024fasta}.

Foldseek was developed to tackle this computational bottleneck, allowing fast structural alignments across thousands of protein structures via the implementation of a 3D interaction alphabet, allowing tokenization of the structural information of the protein backbone. In combination with the previously developed MMseqs2 sequence search software, Foldseek allows fast comparison of thousands of protein structures \parencite{steinegger_clustering_2018,steinegger_mmseqs2_2017,vankempen2024fasta}.

\subsubsection{Structure-based Drug Design}

Recent advances in structure-based drug design have leveraged diverse \ac{ML} frameworks, each offering distinct strengths and methodologies for molecular design. 
\textbf{DeepBlock} employs a \ac{DL} approach that decomposes molecules into synthetically accessible and chemically reactive building blocks, enabling targeted ligand synthesis tailored to protein sequences without requiring explicit 3D structural input \parencite{li2024deep}. 
Its dual-stage framework integrates a conditional variational autoencoder with a graph-based reconstruction algorithm, uniquely combining biochemical rationale and generative precision.

In contrast, \textbf{DiffSBDD} utilizes a 3D SE(3)-equivariant diffusion model, directly generating ligand structures conditioned on protein pocket geometries \parencite{schneuing2024structurebased}. 
It employs symmetry-aware \acp{GNN}, ensuring chemically valid and geometrically consistent outputs, and provides versatile generative capabilities adaptable to various design scenarios without retraining.

Similarly, \textbf{Pocket2Mol} is an E(3)-equivariant generative model that leverages vector-based neural representations and geometric vector perceptrons to autoregressively sample chemically realistic molecules from pocket geometries \parencite{peng2022pocket2mol}. 
This method uniquely balances structural accuracy, generative efficiency, and chemical validity, effectively integrating spatial and chemical bonding constraints.

Further enhancing structural fidelity, \textbf{ResGen} integrates parallel multiscale modeling within a dual-level autoregressive architecture, capturing both local atomic details and global protein--ligand interactions \parencite{zhang_resgen_2023}. 
Its SE(3)-equivariant framework ensures physical plausibility while significantly accelerating the generation of drug-like molecules specifically tailored to protein pockets.

In a different methodological direction, \textbf{TamGen} applies a \ac{GPT}-like chemical language model combined with transformer-based protein encoders to efficiently generate \ac{SMILES}-based compounds from 3D protein pocket data \parencite{wu_tamgen_2024}. 
Its architecture supports both \textit{de novo} molecule generation and iterative refinement without relying on explicit 3D ligand structures during training, enhancing chemical validity, synthetic accessibility, and generation speed.

Lastly, \textbf{TargetDiff} adopts a non-autoregressive, SE(3)-equivariant diffusion model that jointly models atomic coordinates and types, ensuring efficient and realistic molecule generation conditioned on protein binding sites \parencite{guan20233d}. 
Distinctively, it doubles as an unsupervised feature extractor for binding affinity prediction, integrating molecule generation and evaluation within a unified end-to-end framework.

Overall, these models illustrate diverse yet complementary strategies, varying primarily in their reliance on structural versus sequence data, generative methodologies (autoencoder, autoregressive, diffusion-based), and specific design objectives (efficiency, flexibility, biochemical rationale).

\section{Results and Discussion}

Our goal was to design antibiotics that target novel bacterial structures.
To accomplish this, we developed a three-step pipeline:
First, we \textbf{clustered predicted protein structures to identify new antibacterial targets}, leveraging structural alignments to detect conserved protein groups often missed by sequence-based methods.
Next, we \textbf{performed \textit{de novo} molecular design} to generate candidate antibiotics, using rigorous filtering to retain only those with strong structural and biological relevance.
Finally, we \textbf{transitioned from \textit{in silico} predictions to physically realizable compounds} by searching a vast chemical space, enabling rapid and cost-effective synthesis of selected candidates.

\subsection{Target Identification}

As a first step, we clustered the predicted proteomes of seven pathogenic bacterial strains of high medical relevance provided on the AlphaFold database (\cref{sec:struct_clust}) using Foldseek, a software for fast structural clustering of large protein structure databases. The selection contained \textit{E. coli, E. faecum, S. aureus, K. pneumoniae, P. aeruginosa, H. pylori,} and \textit{M. tuberculosis}. The resulting clusters were sorted by the number of essential genes, omitting any cluster containing human analogs. 
Thereby, we identified groups containing genes essential in multiple pathogenic strains, ensuring pan-strain effects, while minimizing toxicity via off-target effects by excluding proteins with homologs in human cells. 
The most promising clusters are shown in \cref{tab:foldseek_cluster}. 
Among our identified targets, multiple had recently been published as novel targets for antibiotic development, such as AccD, FtsW and LolE, with the latter resulting in the novel antibiotic "Lolamicin" with a pan-strain Gram-negative activity. 
These results highlight the potential of our clustering approach, identifying pan-strain antibiotic targets for antibiotics with novel modes of action. Therefore, the clustering approach is a suitable first step of the drug discovery pipeline as it introduces novel targets for application-specific ligand design. 

\begin{figure}[phtb]
    \centering
    \includegraphics[width=\textwidth]{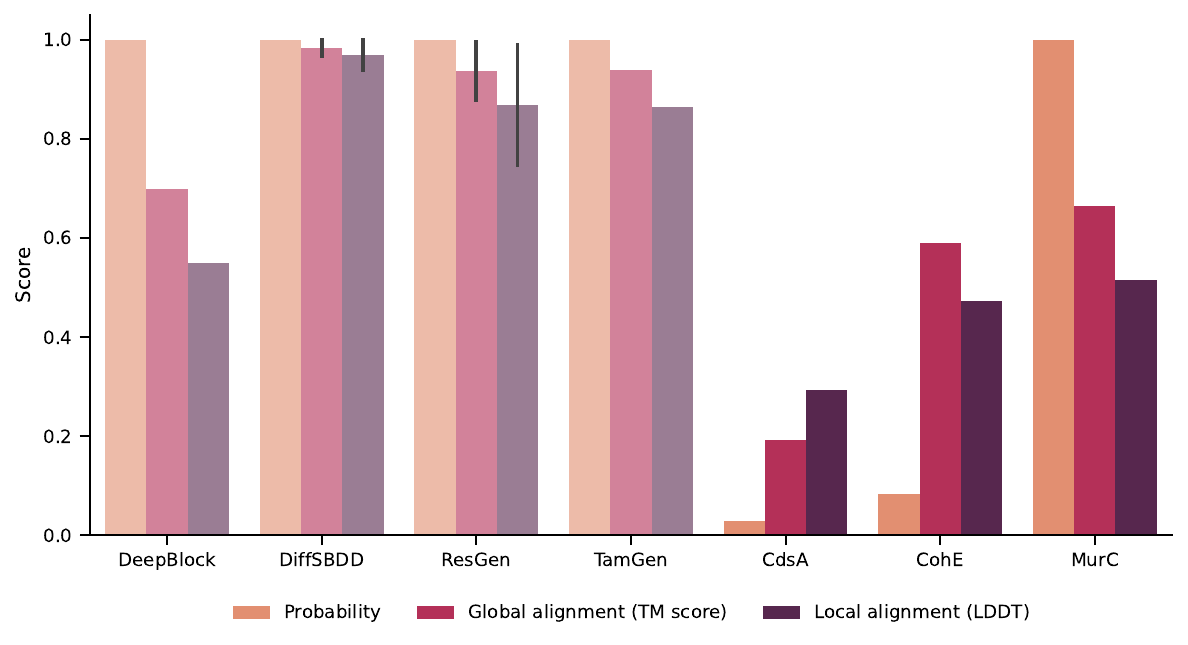}
    \caption{\textbf{Foldseek search similarity of target proteins to CrossDocked2020 training set.} 
    We show the top1 result of each Foldseek search for case study proteins from the respective model publications ($\alpha=0.6$) and our 3 targets ($\alpha=1$) against the CrossDocked2020 training set. 
    For DiffSBDD and ResGen multiple case study proteins were scored. Pocket2Mol and TargetDiff were omitted as the respective publications show now independent case studies. 
    Three metrics are reported: Probability score (Foldseek’s estimated probability of the top1 hit to be homologous to the query structure), the global \ac{TM} score, and \ac{LDDT} score.
    The performed case studies per model were: 
    DeepBlock with NCEH1; 
    DifSBDD with BIKE and MPSK1;
    ResGen with AKT1 and CDK2;
    TamGen with CLPP2\_MYCTU.
    }
    \label{fig:foldseek}
\end{figure}

An important factor during target selection for such a pipeline is the structural similarity of our protein targets to the training set of the subsequently applied generative models. 
Presumably, the closer the targets resemble the training set structures, the better the prediction performance will be. 
All 6 models use the same CrossDocked2020 dataset as thier training set, applying the same sampling and splitting methods. This consistency makes comparisons much easier.
While most of these models show convincing case studies in their respective publications, the protein targets closely resemble structures present in the training sets, as seen in \cref{fig:foldseek}. 

\begin{table}[ptb]
\centering
\caption{\textbf{Essential gene clusters, i.e., potential antibacterial targets.} 
We used Foldseek clustering to identify novel targets. 
All clusters have no close human analogs.
Based on the scientific literature, we investigated whether clusters are targeted and act as enzymes.
Gene names refer to the organism of the representatives.
The cluster size $n$ and the number of essential genes $n_e$ within the cluster are shown.
Our target selection is \colorbox{TUMblue!20}{highlighted}.
}
\label{tab:foldseek_cluster}
\begin{adjustbox}{width=1\textwidth}
\begin{tabular}{lSSp{7.5cm}lll}
	\toprule
	\textbf{Representative} & $\mathbf{n}$ & $\mathbf{n_e} \downarrow$ & \textbf{Protein name} & \textbf{Gene name} & \textbf{Targeted} & \textbf{Catalytic} \\
	\midrule
	\rowcolor{TUMblue!20} Q2FXJ0 & 28 & 13 & UDP-N-acetylmuramate--L-alanine ligase & murC & True & True \\
	
	Q2G167 & 206 & 11 & Putative hemin import ATP-binding protein HrtA & hrtA & False & False \\
	
	P0AFI2 & 11 & 6 & DNA topoisomerase 4 subunit A & parC & True & True \\
	
	O25604 & 12 & 6 & Acetyl-coenzyme A carboxylase carboxyl transferase subunit beta & accD & True & True \\
	
	Q2FYS5 & 11 & 6 & DNA topoisomerase 4 subunit B & parE & True & True \\
	
	\rowcolor{TUMblue!20} P75974 & 12 & 5 & Prophage repressor CohE & cohE & False & False \\
	
	Q9HW01 & 33 & 5 & UDP-N-acetylglucosamine--N-acetylmuramyl-(pentapeptide) pyrophosphoryl-undecaprenol N-acetylglucosamine transferase & murG & True & True \\
	
	Q2FXY3 & 13 & 5 & Probable nicotinate-nucleotide adenylyltransferase & nadD & True & True \\
	
	O33877 & 11 & 5 & 3-hydroxydecanoyl-[acyl-carrier-protein] dehydratase & fabA & True & True \\
	
	P56042 & 6 & 4 & Large ribosomal subunit protein bL17 & rplQ & --- & False \\
	
	A0A133CSM8 & 13 & 4 & UDP-N-acetylglucosamine 1-carboxyvinyltransferase & murA & True & False \\
	
	P75958 & 21 & 4 & Lipoprotein-releasing system transmembrane protein LolE & lolE & True & False \\
	
	A0A0H3GRH8 & 7 & 4 & Outer-membrane lipoprotein LolB & lolB & True & False \\
	
	O25719 & 6 & 4 & Riboflavin biosynthesis protein & --- & True & True \\
	
	Q842S4 & 8 & 4 & Peptide deformylase & def & True & True \\
	
	A0A1A7SPR9 & 19 & 4 & HAD family hydrolase & --- & False & False \\
	
	P66572 & 6 & 4 & Small ribosomal subunit protein uS5 & rpsE & --- & False \\
	
	Q2G2C2 & 11 & 4 & Probable peptidoglycan glycosyltransferase FtsW & ftsW & True & True \\
	
	Q9HWE4 & 12 & 4 & Small ribosomal subunit protein uS17 & rpsQ & --- & False \\
	
	A0A0H3H383 & 8 & 4 & 7,8-dihydroneopterin aldolase & folB & True & True \\
	
	O26088 & 8 & 4 & Organic solvent tolerance-like N-terminal domain-containing protein & --- & False & False \\
	
	A0A133CQC3 & 7 & 4 & CDP-diacylglycerol--glycerol-3-phosphate 3-phosphatidyltransferase & pgsA & True & True \\
	
	Q2FW07 & 6 & 4 & Large ribosomal subunit protein uL4 & rplD & --- & False \\
	
	Q9HXH5 & 8 & 4 & LPS export ABC transporter permease LptG & lptG & True & False \\
	
	P48940 & 6 & 4 & Small ribosomal subunit protein uS7 & rpsG & --- & False \\
	
	A0A132ZIN5 & 10 & 3 & --- & - & --- & - \\
	
	O25916 & 10 & 3 & Replicative DNA helicase DnaB & dnaB & True & True \\
	
	Q2G268 & 5 & 3 & Coenzyme A biosynthesis bifunctional protein CoaBC & coaBC & True & True \\
	
	Q9HUL6 & 6 & 3 & tRNA threonylcarbamoyladenosine biosynthesis protein TsaE & tsaE & False & False \\
	
	Q9HXJ8 & 6 & 3 & GTPase Der & der / yhbZ & True & False \\
	
	A0A132P515 & 5 & 3 & Large ribosomal subunit protein bL20 & rplT & --- & False \\
	
	Q2FYR1 & 5 & 3 & Aminoacyltransferase FemB & femB & True & True \\
	
	A0A132Z939 & 6 & 3 & Chromosomal replication initiator protein DnaA & dnaA & True & False \\
	
	P0AFG0 & 7 & 3 & Transcription termination/antitermination protein NusG & nusG & True & False \\
	
	Q9F1K0 & 8 & 3 & DNA polymerase III subunit alpha & dnaE & True & True \\
	
	A0A0H3GZA8 & 6 & 3 & Large ribosomal subunit protein uL22 & rplV & --- & False \\
	
	\rowcolor{TUMblue!20} P76091 & 8 & 3 & Uncharacterized protein YnbB & ynbB & False & True \\
	
	P56038 & 6 & 3 & Large ribosomal subunit protein uL13 & rplM & --- & False \\
	
	P56031 & 5 & 3 & Large ribosomal subunit protein uL3 & rplC & --- & False \\
	
	P56036 & 6 & 3 & Large ribosomal subunit protein uL10 & rplJ & --- & False \\
	
	P15042 & 8 & 3 & DNA ligase & ligA & True & True \\
	
	P06710 & 5 & 3 & DNA polymerase III subunit tau & dnaX & False & True \\
	
	O25688 & 8 & 3 & Ribosome-binding factor A & rbfA & --- & False \\
	
	P56102 & 9 & 3 & Methionine aminopeptidase & map & True & True \\
	
	P08373 & 5 & 3 & UDP-N-acetylenolpyruvoylglucosamine reductase & murB & True & True \\
	
	P52097 & 5 & 3 & tRNA(Ile)-lysidine synthase & tilS & True & True \\
	
	A0A1X4JGB4 & 5 & 3 & tRNA (Adenosine(37)-N6)-threonylcarbamoyltransferase complex dimerization subunit type 1 TsaB & tsaB & False & False \\
	
	Q9HWE2 & 6 & 3 & Large ribosomal subunit protein uL16 & rplP & --- & False \\
	
	A0A133CHI6 & 11 & 3 & Acyl carrier protein & acpP & --- & False \\
	
	A0A133CVC2 & 5 & 3 & Translation initiation factor IF-3 & infC & True & False \\
	\bottomrule
\end{tabular}
\end{adjustbox}
\end{table}

To explore how structural similarity to the training data influences model performance, we selected three targets with varying degrees of resemblance to the training set: MurC, CdsA, and CohE. 
MurC was chosen as a "positive control," as it has already been shown to be a valid antibiotic target with a published inhibitor \parencite{humnabadkar_udp-_2014}, and it has a close structural analogue in the CrossDocked2020 training set.  
CdsA was selected as a more challenging candidate, as it is a novel enzymatic target with no known inhibitors, and structurally distant from the training set (\cref{fig:foldseek}). 
However, it has a probable binding pocket for its enzymatic activity, according to AlphaFold 3 predictions using its native ligand, making the inhibitor prediction more straightforward. 
Finally, CohE was picked as a high-risk target, as it is a putative phage repressor without clear enzymatic activity and no close analogs in the training set. 
Unlike the other targets, we aim to disprupt its potential dimerization by binding into a putative protein-protein dimerization interface identified via AlphaFold, which is hinted at by the structurally similar known transcription factor LexA \parencite{luo_crystal_2001}. 
As neither the dimerization interface nor the structure or function of CohE were investigated thus far, it is a highly interesting yet highly speculative third target.

\subsection{Molecule Generation}

The focal point of this study is the comparison of different \ac{SBDD} methods and their applicability in real world drug discovery settings. 
The models used in this study were chosen based on the corresponding publication, code- and pretrained weights availability, and underlying model archetype to cover a broad selection of model types.
Consequently, the following models were evaluated: DeepBlock, DiffSBDD, Pocket2Mol, ResGen, TamGen, and TargetDiff \parencite{li2024deep,guan20233d,peng2022pocket2mol,zhang_resgen_2023,wu_tamgen_2024,schneuing2024structurebased}.

\begin{figure}[phtb]
    \centering
    \includegraphics[width=\textwidth]{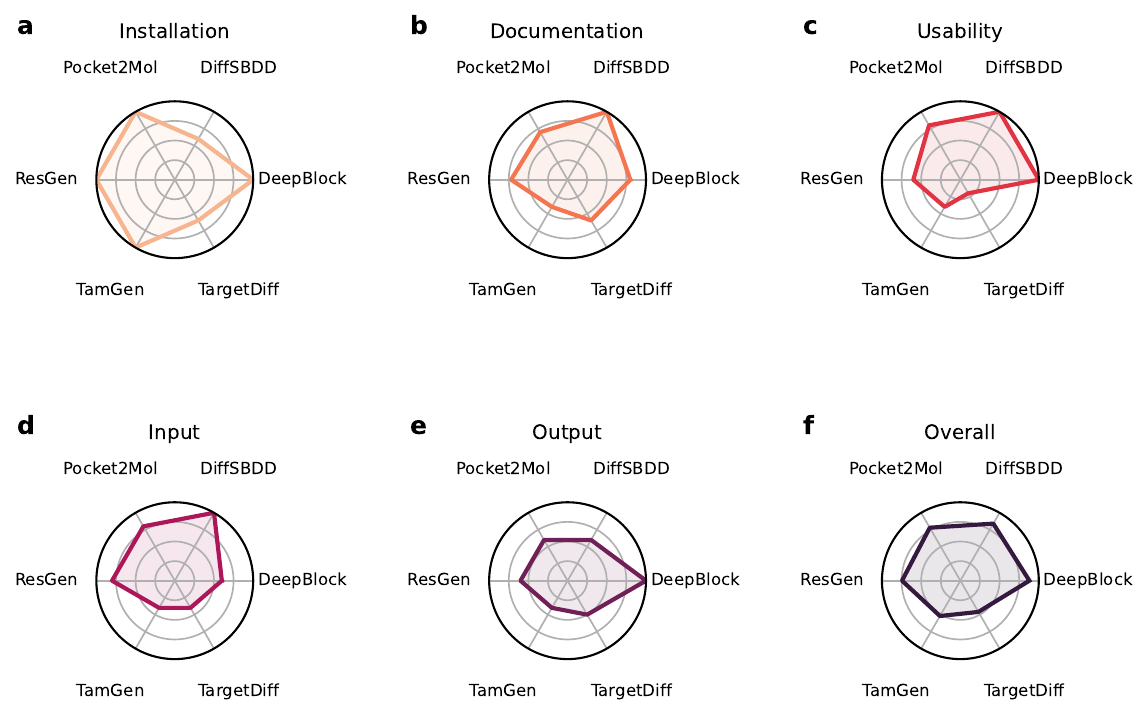}
    \caption{\textbf{Model implementation ranking.} 
    \textbf{a)} We ranked the ease of installation from the respective GitHub repositories.
    \textbf{b)} We rated the documentation of the repositories and the underlying codebases.
    \textbf{c)} We compared the models' usability, focused on sampling with pretrained models.
    \textbf{d)} We rated the input options and ease of input generation.
    \textbf{e)} We evaluated the output format, especially the contents and ease of further data processing.
    \textbf{f)} We showed an overall rating averaged across the previous characteristics.}
    \label{fig:implementation}
\end{figure}

\paragraph{Implementation}
One important factor for researchers interested in integrating novel \ac{ML} models into their pipeline is the ease of implementation, especially for those with limited computational experience. 
Therefore, as a first step, we sought to compare the implementation and deployment of the respective GitHub repositories (\cref{fig:implementation}), omitting any web servers, as proper integration into drug discovery pipelines usually necessitates local model implementations. 
It should be noted that the ranking reflects, to a certain extent, our personal experience with the implementation and use of the methods being compared.

All models supply some kind of virtual environment file, Dockerfile or similar method of setup, making installation straightforward. 
For DiffSBDD and TargetDiff, we experienced issues with the installation environments, resulting in manual solutions to package version issues. 
However, some of these issues could be hardware-specific, so installation experience might be setup-dependent.

The quality of documentation varied strongly between repositories, with some, e.g., DiffSBDD, having extensive documentation with well described step-by-step guides on their main repository page. 
Others were less straightforward and required more in-depth understanding of the different scripts and necessary preprocessing, such as TamGen.

Usability is probably the most important metric in this section, and shows clear differences amongst the models. 
Some, e.g., DiffSBDD and DeepBlock, are straightforward and do not require the user to look any further than the main repository page, only needing a single line of code to execute the entire predictions. 
Other models, such as TamGen and TargetDiff, required extensive preprocessing. 
TargetDiff, for example, required elaborate pocket extraction before being able to run, while TamGen required the user to first generate a TamGen-dataset before running the actual ligand generation script. 
Both these preprocessing steps were not straightforward due to limited documentation.

Most models require some sort of pocket input, either straightforward coordinates or amino acid indices, or more complex input in the form of preprocessed datasets or extracted pockets. 
DeepBlock, unlike the other models, does not allow for pocket selection and instead uses an intrinsic binding site detection. 
This might be beneficial for ligand generation for targets without a known binding site. 
However, as we focus on site-specific ligand generation, this aspect limited DeepBlock's applicability compared to the other models.

Finally, the output was scored regarding usability in subsequent data analysis. 
Most models generated simple \texttt{.sdf} files or \ac{SMILES} strings. 
DeepBlock outputs additional information like drug-likeliness and logP values, which distinguishes it from other models.

Overall, DeepBlock performed best regarding implementation and usability, despite the inability to specify a binding site. 
TamGen and TargetDiff performed worst overall, mostly due to elaborate and inadequately documented preprocessing steps. 
This is, however, a subjective evaluation of our experience, and, more importantly, is no verdict regarding actual molecule design performance.

\paragraph{Generation}

\begin{figure}[phtb]
    \centering
    \includegraphics[width=\textwidth]{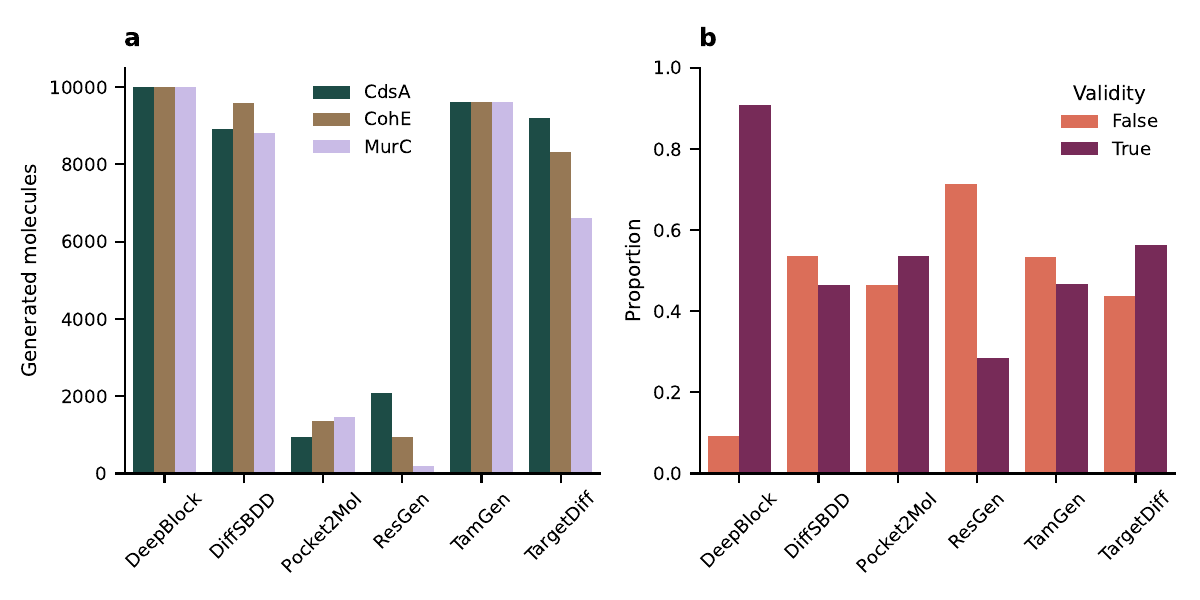}
    \caption{\textbf{Overview of generated molecules.} 
    \textbf{a)} We visualize the number of actually generated molecules based on model and target protein. 
    We prompted all models to generate \num{10000} molecules per target.
    \textbf{b)} We show the validity of molecules in relation to the \ac{SBDD} model. 
    The validity is determined by our \texttt{SMILESCleaner} class.}
    \label{fig:gen_overview}
\end{figure}

We aimed to generate \num{10000} molecules per target and model. 
However, not all models achieved this mark, particularly Pocket2Mol and ResGen (\cref{fig:gen_overview}a).
This limitation most likely stems from the models' internal checks for duplicity and validity. 
If a model exceeds a certain threshold of failed attempts, it exits the generation loop. 
The likely cause of not reaching the desired number of designs is poor generalization, which causes the model to get stuck on certain motifs and, in turn, failing to explore a broader chemical space.
TamGen could only generate 1000 molecules, so as suggested for LLMs, we used 10 seeds to generate the desired number of structures.
Overall, we generated \num{117231} molecules across all methods.

Next, we computed the individual percentage of valid molecules generated by each model using our \texttt{SMILESCleaner} (\cref{sec:SMILESCleaner}), which checks for validity of \ac{SMILES}. 
This step was necessary because internal cleaning methods vary across models, depending on their implementation and architecture.
DeepBlock outperformed all other models, with \SI{\sim90}{\percent} of its generated molecules being valid. 
Most other models achieved around \SI{50}{\percent} validity, except ResGen, which produced only \SI{\sim30}{\percent} valid molecules (\cref{fig:gen_overview}b). 
In total, out of the original \num{117231} molecules, \num{69867} were found to be valid across all models.

\begin{figure}[phtb]
    \centering
    \includegraphics[width=\textwidth]{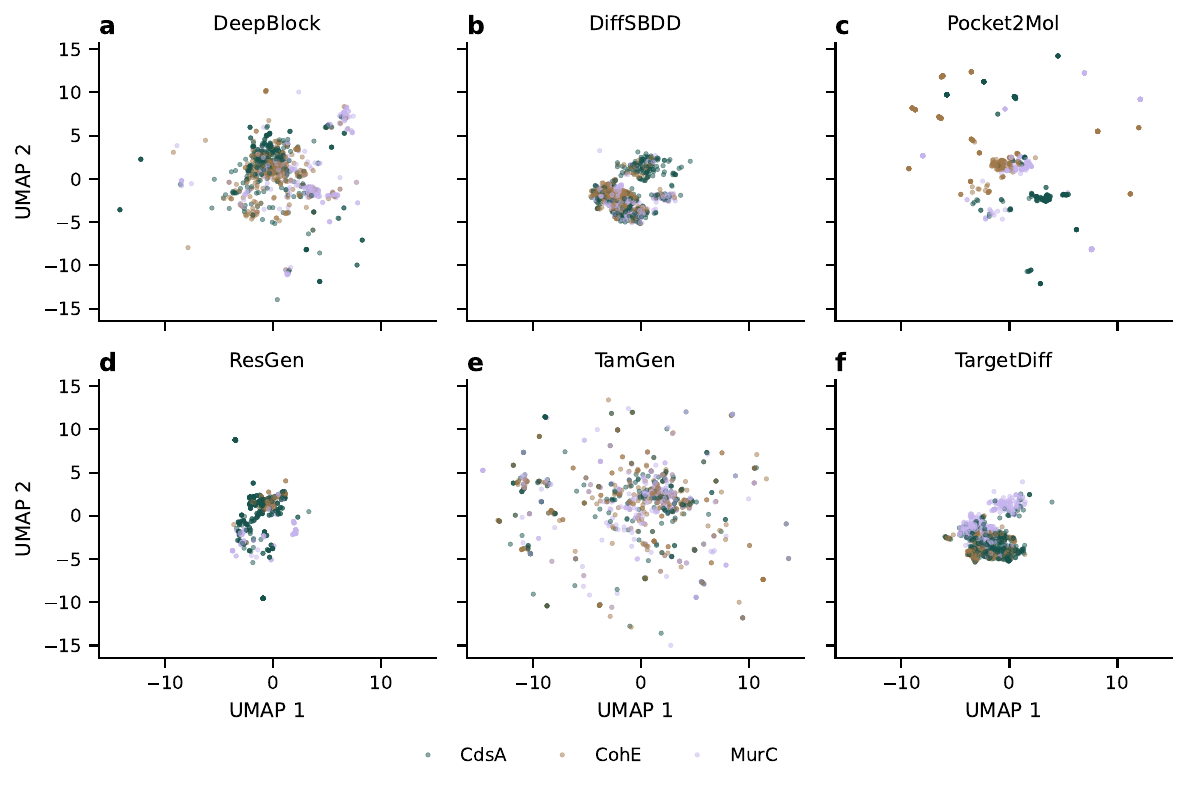}
    \caption{\textbf{Model-based chemical space analysis of the generated molecules.} 
    The UMAP space is presented for each model separately (\textbf{a--f}), based on ECFPs.
    Colors represent the different protein targets.
    We sampled up to 1000 molecules per subplot. 
    Only valid molecules were considered.}
    \label{fig:umap_model}
\end{figure}

We also analyzed the chemical space and diversity of the generated molecules (\cref{fig:umap_model,fig:umap_target}).
TamGen probed the chemical space the most, while DeepBlock and Pocket2Mol balanced between exploring and exploiting scaffold diversity.
In contrast, DiffSBDD, ResGen, and TargetDiff produced structurally less diverse molecules compared to other methods and showed pronounced clusters. 

An interesting finding is that TamGen generates the same scaffolds, regardless of the target, in approximately \SI{10}{\percent} of all \textit{de novo} generations (\cref{fig:venn_models_grid}).
For CdsA and CohE, \num{49751} molecules were generated, of which \num{4098} were redundant and target-agnostic---significantly more than DeepBlock, the model with the next highest redundancy, with only \num{613} such molecules. 
A similar trend was observed for other target pairs: out of \num{45800} molecules generated by TamGen for CdsA and MurC, \num{4478} were target-agnostic, while for CohE and MurC, \num{3821} out of \num{44183} molecules showed the same redundancy.
This behavior likely stems from TamGen's architecture, as it is the only LLM-based method. 
Repeated generation of the same molecule suggests low perplexity toward this scaffold, which can be interpreted as the model's "confidence" in it. 
However, it also indicates that TamGen may sometimes lack deep structural awareness.

\paragraph{Curation}
Our goal is to efficiently identify promising scaffolds among \num{69867} candidates, a challenge that can be framed as a "needle in a haystack" problem.
To narrow the search, we applied filters commonly used in medicinal chemistry and antibacterial drug discovery. These include the \ac{REOS} filters, which attempt to enrich compounds with desirable or "drug-like" properties \parencite{walters_virtual_1998,walters1999recognizing}.

Finding new antibiotics is particularly challenging because the chemical space traditionally explored for drug discovery---spanning academic, institutional, and commercial libraries---differs significantly from the space occupied by known antibiotics.
To address this, researchers have defined a distinct property space that characterizes most antibiotics \parencite{reck2019challenges,miethke2021sustainable}.

\begin{figure}[phtb]
    \centering
    \includegraphics[width=\textwidth]{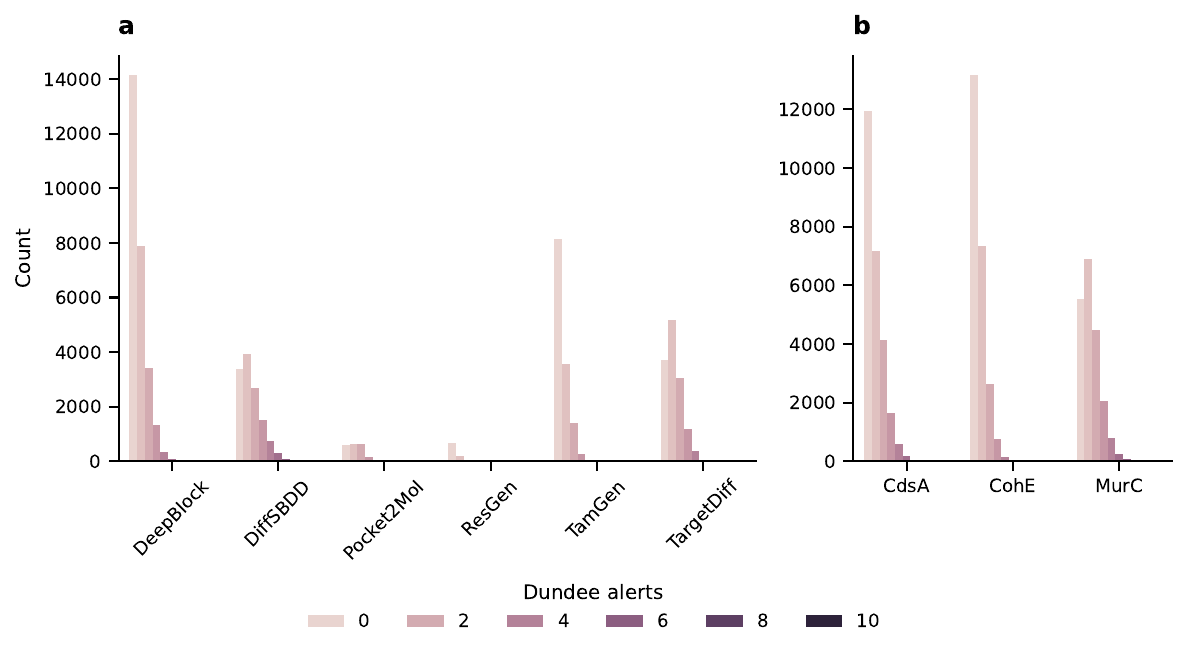}
    \caption{\textbf{Structural alerts of generated molecules.} 
    Here, we analysed the number of Dundee alerts based on model (\textbf{a}) as well as target (\textbf{b}). 
    Only valid molecules were considered. 
    }
    \label{fig:warnings_overview}
\end{figure}

\begin{figure}[phtb]
    \centering
    \includegraphics[width=\textwidth]{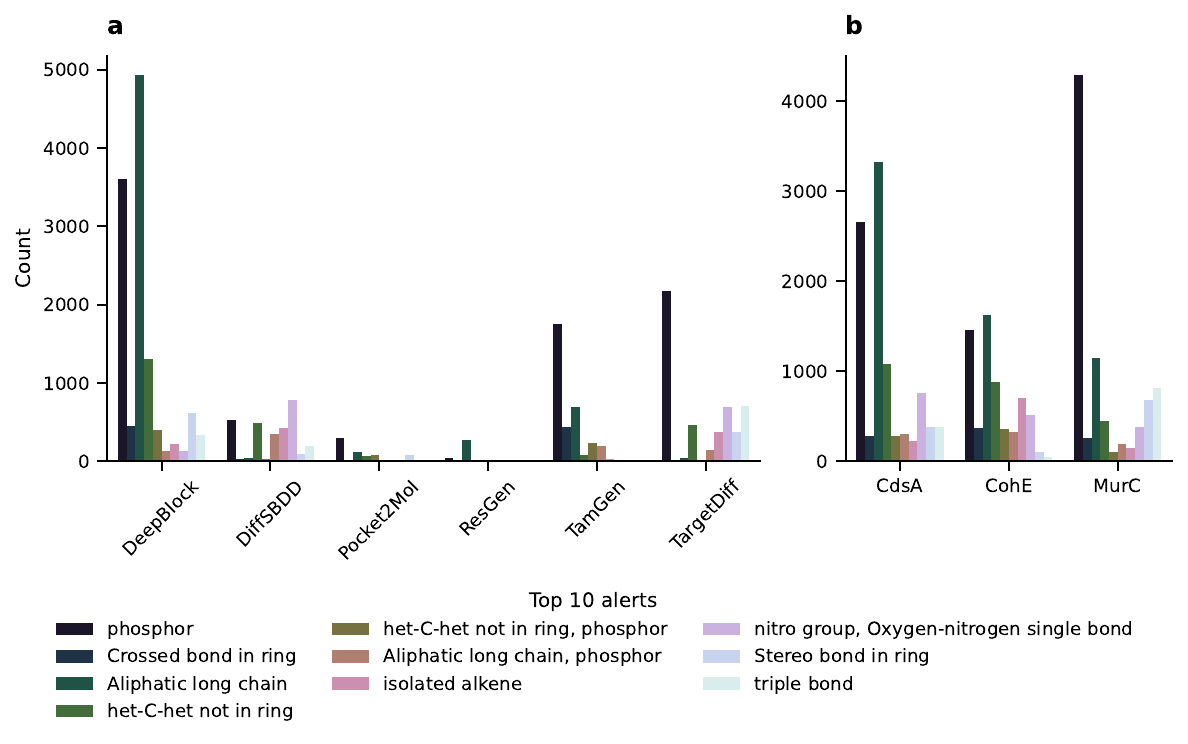}
    \caption{\textbf{Top ranking structural alerts of generated molecules.} 
    Here, we investigated which Dundee alerts are the most frequent based on model (\textbf{a}) as well as target (\textbf{b}). 
    Only valid molecules were considered.}
    \label{fig:error_analysis_combined}
\end{figure}

First, we analyzed the number of Dundee alerts generated in relation to both the model and the target (\cref{fig:warnings_overview}), a subset of the \ac{REOS} structural alert library. These libraries are derived from medicinal chemistry rule sets to filter for compounds with high drug-likeliness.
The models DeepBlock and TamGen generated more molecules with fewer \ac{REOS}/Dundee alerts compared to all other methods.
Notably, the distribution at the target level reveals that MurC as a target input resulted in less drug-like molecules than the other targets.
We further investigated the underlying causes of these alerts (\cref{fig:error_analysis_combined}).
The main contributors were phosphor-containing compounds, spiro ring systems, and long aliphatic chains. 
These appear to be molecular artefacts learned during training---structures that are frequently introduced but generally undesired in drug discovery. However, some alerts might be too restrictive and can be altered based on the users application context.

\begin{figure}[phtb]
    \centering
    \includegraphics[width=\textwidth]{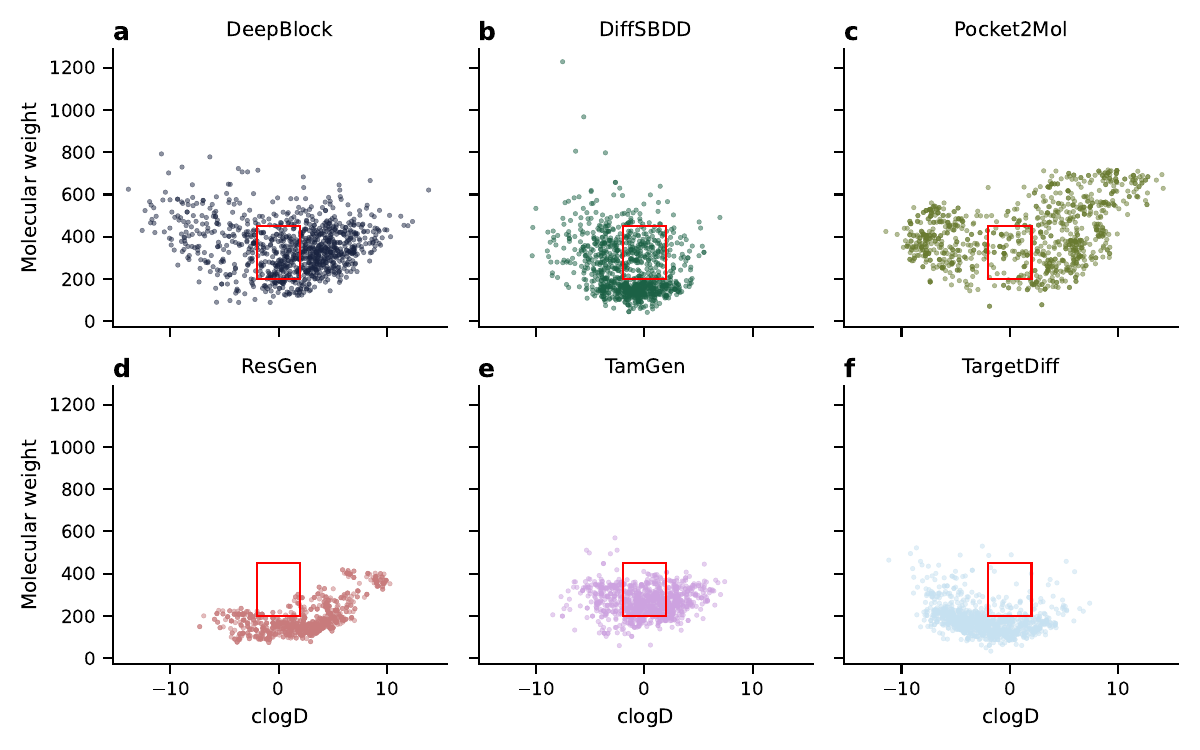}
    \caption{\textbf{Property space of generated molecules.} 
    The space is presented separately for each model (\textbf{a--f}). 
    Known cytoplasmic antibiotics property space is marked as a \textcolor{red}{box}. 
    We visualized molecular weight and calculated logD at pH 7.4. 
    We sampled up to 1000 molecules per subplot. 
    Only valid molecules were considered.}
    \label{fig:scatter_models_grid}
\end{figure}

Second, we refined the desired output chemical space using physiochemical principles specific to antibiotic design. 
Corporate compound libraries tend to cluster around higher molecular weight and lipophilicity due to legacy synthesis strategies. 
Unlike other drug classes, antibiotics exhibit distinct property profiles, allowing us to tailor our search for novel antibacterial scaffolds. 
Analysis of existing antibiotics emphasizes the relevance of target location in both hit discovery and lead optimization. 
Given that our targets reside in the cytoplasm, we prioritized physiochemical properties that enhance intracellular accumulation. 
These preferences align with patterns reported across antibiotic discovery efforts \parencite{reck2019challenges,brown2014trends,tommasi2015eskapeing}. 
Accordingly, we selected compounds with clogD values between \numrange{-2}{2} and molecular weights from \SIrange{200}{450}{\dalton}.
\Cref{fig:scatter_models_grid} illustrates this narrowed selection space (target and alert-based investigations can be found in \cref{fig:scatter_target_dundee}).
One key observation is that the models---Pocket2Mol, ResGen, and TargetDiff---produced few compounds within the desired property window. Since all methods were trained on the same dataset, this discrepancy likely stems from differences in model architecture. 

\subsection{Realization}
After testing for validity and drug-likeliness using the REOS/Dundee alters we ended up with \num{30662} molecules for all three targets.
Before applying the physicochemical filters, we sought to assess the number of compounds we could reasonably acquire to ensure the general accessibility of the approach.
Chemical vendors provide a vast chemical space as hypothetical compounds, which can be synthesized through automated and fragment-based methods (\cref{tab:spaces}). 
However, screening these enormous spaces creates an even greater bottleneck, akin to finding a "needle in a \emph{needle}stack."
Luckily, tools like SpaceMACS enable users to find close analogs of their compounds of interest among trillions of hypothetical molecules \parencite{schmidt2022maximum}.

\begin{figure}[phtb]
    \centering
    \includegraphics[width=\textwidth]{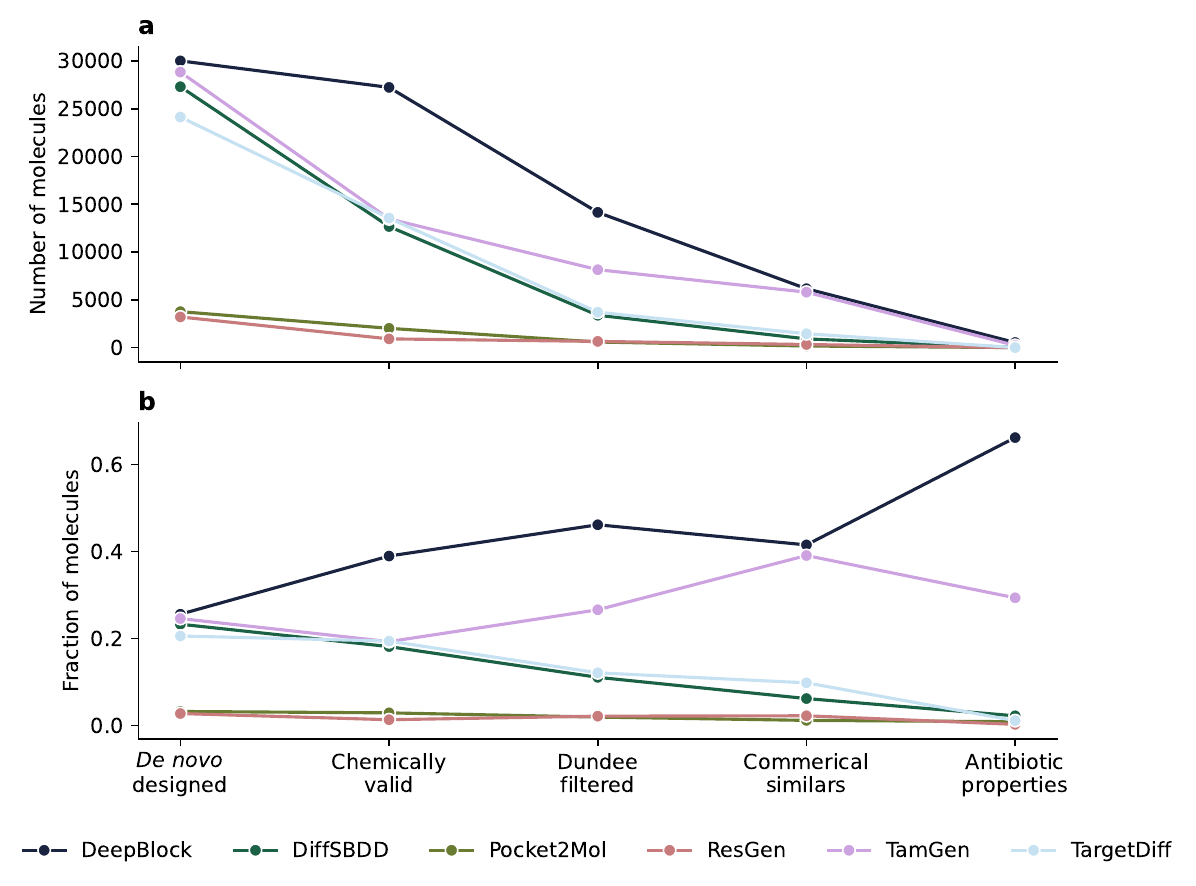}
    \caption{\textbf{Processing of \textit{de novo} structures.} 
    We evaluated all models according to the curation and filtering procedure. 
    First, we checked the validity of all generated molecules for the design campaigns against the three targets. 
    Next, we removed any molecules that triggered \ac{REOS}/Dundee alerts. 
    Then, we assessed how many of the generated scaffolds could be commercially acquired, requiring a \acs{MCS} similarity $\geq 0.9$ to the generated structures. 
    Finally, we applied an additional filter to select molecules that meet the physicochemical requirements associated with antibiotic activity.
    This is expressed both as the absolute number of molecules remaining at each step (\textbf{a}) and as the fraction relative to all molecules retained at that processing stage (\textbf{b}).}
    \label{fig:molecules_pipeline}
\end{figure}

To identify molecules suitable for laboratory testing, we followed a systematic multi-step curation process (\cref{fig:molecules_pipeline}). 
We began by searching for commercially available analogs of all valid, Dundee-filtered molecules, using SpaceMACS to screen the commercial spaces referenced in \cref{tab:spaces}. 
From this set, we retained only those analogs with a high structural similarity, defined by a $\text{\acf{MCS} similarity} \geq 0.9$. 
To further refine the selection, we chose a single representative per Murcko scaffold---the molecule with the highest \ac{MCS} score. 
Physiochemical property filters regarding molecular weight and calculated logD were then applied to yield our final candidate set. 
Most molecules that passed this full pipeline originated from DeepBlock and TamGen, while none of the compounds generated by TargetDiff advanced beyond the curation or analogue search stages. 

\begin{figure}[phtb]
    \centering
    \includegraphics[width=\textwidth]{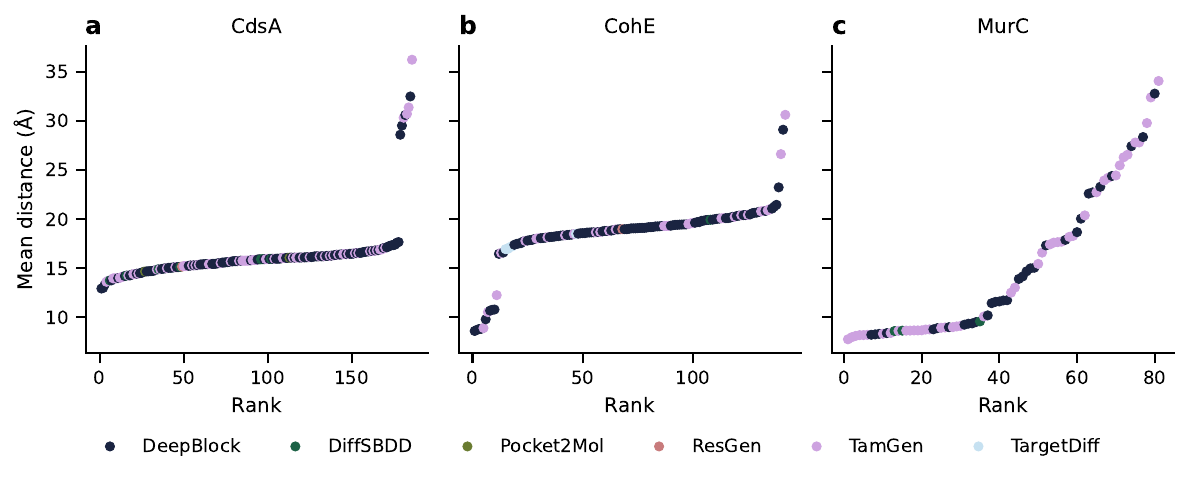}
    \caption{\textbf{Distance to pocket based on complex structures.} 
    We predicted complex structures for all selected molecule–target pairs using AlphaFold 3. 
    Then, all molecules were ranked by mean distance.
    For each protein target (\textbf{a--c}), we then calculated the mean distance between the bound ligand and the pocket center, as defined during the \ac{SBDD} setup.}
    \label{fig:af3_ranking}
\end{figure}

Subsequently, we used AlphaFold 3 to predict complex structures comprising the designed ligands and their respective targets, selecting only those ligands positioned near the intended binding pocket or surface (\cref{fig:af3_ranking}).
Upon closer examination of the mean distance, we observed sharp transitions---referred to as cliffs. 
These cliffs arise from variations in the pocket sizes of each target: as long as molecules are predicted to bind within the defined pocket region, the distance remains relatively stable (i.e., plateaus). 
However, as soon as binding occurs outside this region, the distance increases abruptly.
We will only proceed with molecules that are on the plateaus in front of the "cliff edge" in order to find true binders, i.e. to avoid binders that do not show the desired physiological effects.

\section{Conclusions and Outlook}

The goal of this study was twofold: first, to identify new targets for antibacterial treatment, and second, to design specific binders that would inhibit the target function.
The study culminated in a systematic comparison of leading \ac{SBDD} models within a practical drug discovery pipeline.

As the first step, we curated a set of novel antibiotic targets to serve as representative case studies for model benchmarking. 
We then performed a comparative evaluation of the models from an implementation and usability standpoint---acknowledging the subjective nature of this work---with the goal of informing and guiding future tool development. 
In this context, DeepBlock and DiffSBDD showed clear advantages over other methods, especially in terms of preprocessing usability and documentation quality.

The focus of our analysis, however, was on the predictive performance of the models. 
Significant differences emerged in key metrics such as molecular validity, structural diversity, drug similarity, and target specificity. 
DeepBlock consistently outperformed its counterparts, yielding the highest proportion of chemically valid, structurally diverse, and drug-like compounds.

Strikingly, after rigorous post-generation filtering, we found that \SI{61}{\percent} of all commercially available candidate molecules originated from DeepBlock, with TamGen contributing \SI{34}{\percent} and the remaining models accounting for only \SI{5}{\percent}.

We now procure a subset of these compounds for each target, with plans to evaluate antibiotic activity, confirm target engagement and investigate cellular uptake properties.

\section*{Code and Data Availability}

Once all the data is collected, we plan to release all the data and code under the FAIR principles.

\begin{ack}
The authors thank Merck KGaA Darmstadt for their generous support with the Merck Future Insight Prize 2020. 
This project is also cofunded by the European Union (ERC, breakingBAC, 101096911).

We would also like to thank Dr. Thomas Schlichthärle, Prof. Martin Zacharias, and Prof. Martin Steinegger for the many fruitful discussions and suggestions.
Further, we thank Anna M. Pietschmann and Aleksandra Daniluk for their valuable feedback on the manuscript.
\end{ack}

\printbibliography

\appendix

\section{Methods}

\subsection{Structural Clustering Analysis}\label{sec:struct_clust}

We installed the Foldseek (commit \texttt{427df8a}) Conda environment to facilitate the computational workflow.
To identify structurally homologous protein clusters, we used these \texttt{easy-cluster} settings: \texttt{--lddt-threshold 0.5} to accept alignments with a \ac{LDDT} score above the specified threshold, and \texttt{-e 0.1} where high values indicate more distant structural matches.
We selected these values as a balance between sensitivity and specificity \parencite{mifsud2024mapping,vankempen2024fasta}.
A lower threshold increases cross-species alignments, leading to more clusters that contain human homologous, whereas a higher threshold results in fewer clusters due to increased stringency in alignment scoring.

\paragraph{Proteome Data Acquisition}
Proteomes were obtained from the AlphaFold Database Proteome Downloads. 
The following species were selected and retrieved on the respective dates:

\begin{itemize}
    \item \textit{Escherichia coli}, \textbf{reference organism} in our study (November 5th, 2024)
    \item \textit{Enterococcus faecium} (November 6th, 2024)
    \item \textit{Staphylococcus aureus} (November 6th, 2024)
    \item \textit{Klebsiella pneumoniae} (November 6th, 2024)
    \item \textit{Pseudomonas aeruginosa} (November 6th, 2024)
    \item \textit{Helicobacter pylori} (November 7th, 2024)
    \item \textit{Mycobacterium tuberculosis} (November 8th, 2024)
    \item \textit{Homo sapiens} (November 11th, 2024)
\end{itemize}

\paragraph{Data Processing and Essential Gene Selection}
For essential gene identification, clusters (stored in \texttt{*\_cluster.tsv}) containing one or more essential genes were selected, and those including any human proteins were excluded. 
The final dataset consisted of clusters that contained essential bacterial genes with structural homologous in ESKAPE pathogens but without direct human analogs, enabling their potential as antimicrobial targets.

\subsection{Structure-based Drug Design}

We selected binding site of CdsA (modelled using AlphaFold 3 and validated with know substrate binder structures), MurC (annotation in UniProt), and CohE (dimerization surface) residues as model input.
If coordinates were required, they were determined via center of mass in PyMOL of the relevant residues.

\begin{table}[phtb]
\centering
\caption{All \acs{SBDD} models used in this work.}
\label{tab:sbdd_models}
\begin{tabular}{llr}
\toprule
\textbf{Name} & \textbf{Type} & \textbf{GitHub Commit}\\
\midrule
DiffSBDD & Diffusion  & \texttt{de4f605} \\
ResGen & Hierarchical autoregression & \texttt{2db3def} \\
DeepBlock & Building blocks gen., mol. reconstruction & \texttt{444a623} \\
TamGen & Language model & \texttt{9459b3f}\\
TargetDiff & 3D Equivariant Diffusion & \texttt{142f1eb}\\
Pocket2Mol & Equivariant \ac{GNN} & \texttt{836a0c4} \\
\bottomrule
\end{tabular}
\end{table}

\subsubsection{Modifications}
\paragraph{ResGen} We adapted the ResGen code allowing to set a center without ligand, without changing any implementation details or affacting training weights.

\paragraph{TargetDiff}

To define the center of the binding pocket, a pseudoatom was created in PyMOL at the desired spatial coordinates. 
This pseudo-atom was saved as a PDB file and subsequently converted to SDF format using OpenBabel.
An index file was then generated. 
This file contained metadata required for downstream processing, including references to the protein structure (PDB), the pseudo-ligand (SDF), and a placeholder RMSD value. 
Finally, pocket extraction was performed using a Python script. 
This script processed each target protein directory, using the provided center and a fixed radius to generate the final pocket representations.

\paragraph{TamGen}

To prepare the input structures, each target protein file in PDB format was first converted to CIF format. 
This conversion was performed using an online tool such as the PDBx/mmCIF converter available at \url{https://mmcif.pdbj.org/converter/index.php?l=en}.

Next, a dataset file in CSV format was created to define the targets. 
This file included the protein identifier and the three-dimensional coordinates $(x, y, z)$ corresponding to the center of the binding pocket. 
All target proteins were included in this dataset file.

A custom script was then used to generate the dataset directory structure based on the contents of the CSV file and the corresponding CIF protein files. 
This process organized the input files into a consistent format expected by downstream processing tools.

Finally, ligand generation was performed for each protein in the dataset using a beam search-based inference script. 
This procedure was repeated across multiple random seeds to ensure diversity in the generated ligands. 
The resulting ligand candidates were stored in designated output directories for further analysis.

\subsection{Curation} \label{sec:SMILESCleaner}

The \texttt{SMILESCleaner} class was applied and performs the following cleaning steps:

\begin{enumerate}
    \item Tries to parse \ac{SMILES}. (If failed, marks molecules as invalid.)
    \item De-duplicates. (If failed, marks molecules as invalid.)
    \item Cleans \ac{SMILES} using ChEMBL standardizer and returns parent MolBlock. (If failed, marks molecules as invalid.)
    \item Removes all unstable rings. 
    Those are defined as unstable if they do occur less than 50 times is ChEMBL \parencite{walters2024generative}. 
    (If failed, marks molecules as invalid.)
    \item Applies \ac{REOS} filters ("Dundee" set) \parencite{walters1999recognizing}. 
    (We modified the code to not return invalid molecules, but number of \ac{REOS} alerts.)
\end{enumerate}

\subsection{Search for commercially available Compounds}

We selected 6 chemical suppliers (\cref{tab:spaces}).
We identified the most similar using SpaceMACS \parencite{schmidt2022maximum}.
The \ac{MCS} similarity metric favours results with a smaller \ac{MCS} but a comparable size to the query over those with a larger \ac{MCS} but a completely different size.
This is why we chose \ac{MCS} similarity instead of size.
To identify structurally similar compounds, a chemical space search was performed using the SpaceMACS tool. 
For each target, the \ac{SMILES}-derived SDF input file was compared against a molecular space library (\texttt{.space} files). 
The search parameters were configured to return a maximum of five results per query, with a minimum similarity threshold of 0.5. 
The search was executed using 64 computational threads to optimize performance, and the verbosity level was set to 4 for detailed logging. 
Each output was saved as a CSV file named according to the corresponding gene or target.

\begin{table}[h]
    \centering
    \sisetup{scientific-notation = true, round-mode = places, round-precision = 1}
    \caption{Chemical libraries ("Spaces") and their sizes. Total is not the total number of unique molecules because overlap is not disclosed.}
    \label{tab:spaces}
    \begin{tabular}{l S[table-format=1.1e2] l}
        \toprule
        \textbf{Space} & \textbf{Molecules} & \textbf{Provider} \\
        \midrule
        Freedom Space  & 1.4e11  & Chemspace \\
        \textbf{REAL Space}     & 7.0e10  & Enamine \\
        AMBrosia       & 1.1e11  & Ambinter \\
        eXplore        & 5.0e12  & eMolecules \\
        GalaXi         & 1.2e10  & WuXi LabNetwork \\
        CHEMriya       & 1.2e10  & OTAVA \\
        \cmidrule{2-2}\morecmidrules\cmidrule{2-2}
        \textbf{Total} & 5.3e12 & \\
        \bottomrule
    \end{tabular}
\end{table}

\section{Further Results}

\begin{figure}[phtb]
    \centering
    \includegraphics[width=\textwidth]{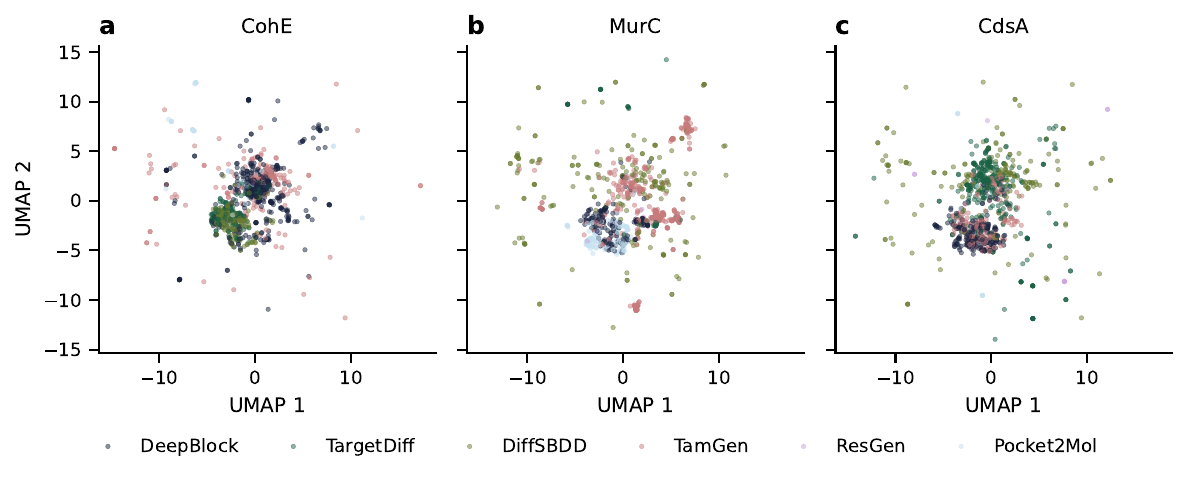}
    \caption{\textbf{Target-based chemical space analysis of generated molecules.} 
    The space is presented as a UMAP on a per target basis (\textbf{a--c}) based on ECFP. 
    Colors represent the different \textit{de novo} model.
    We sampled up to 1000 molecules per subplot. 
    Only valid molecules were considered.}
    \label{fig:umap_target}
\end{figure}

\begin{figure}[phtb]
    \centering
    \includegraphics[width=\textwidth]{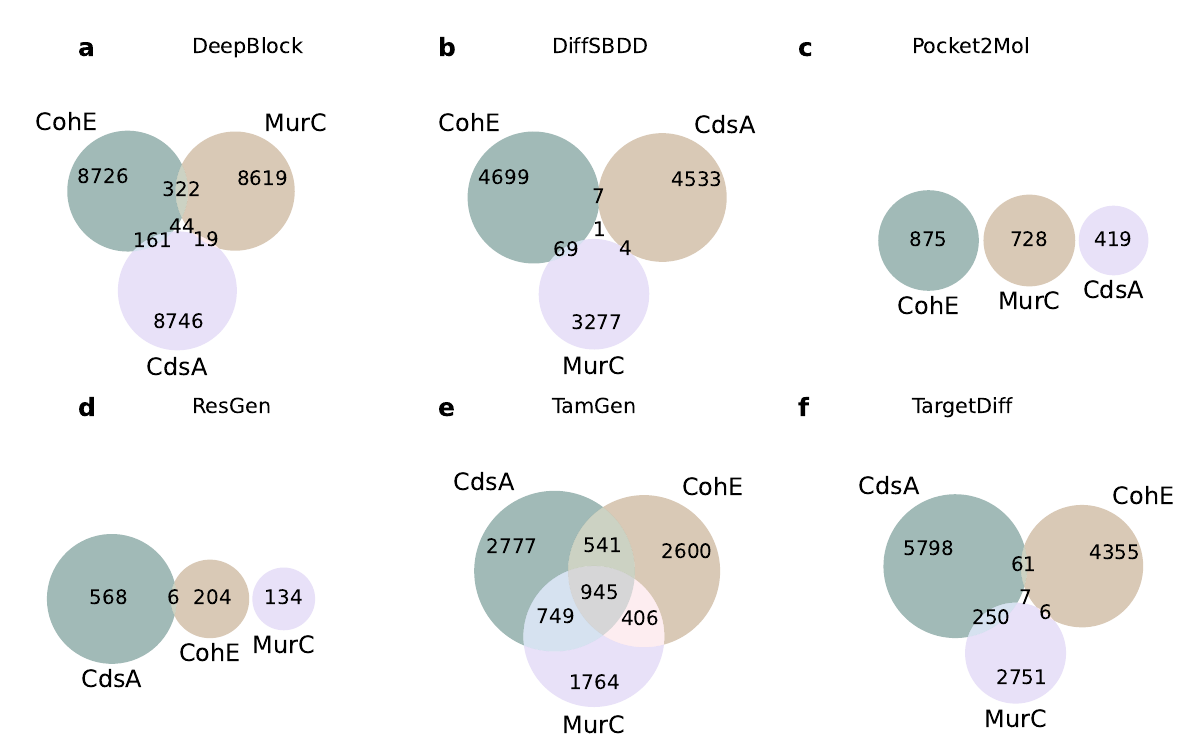}
    \caption{\textbf{Venn diagram of \textit{de novo} molecule redundancy.} 
    Each model (\textbf{a--f}) was checked for target-agnostic molecule predictions.}
    \label{fig:venn_models_grid}
\end{figure}

\begin{figure}[phtb]
    \centering
    \includegraphics[width=\textwidth]{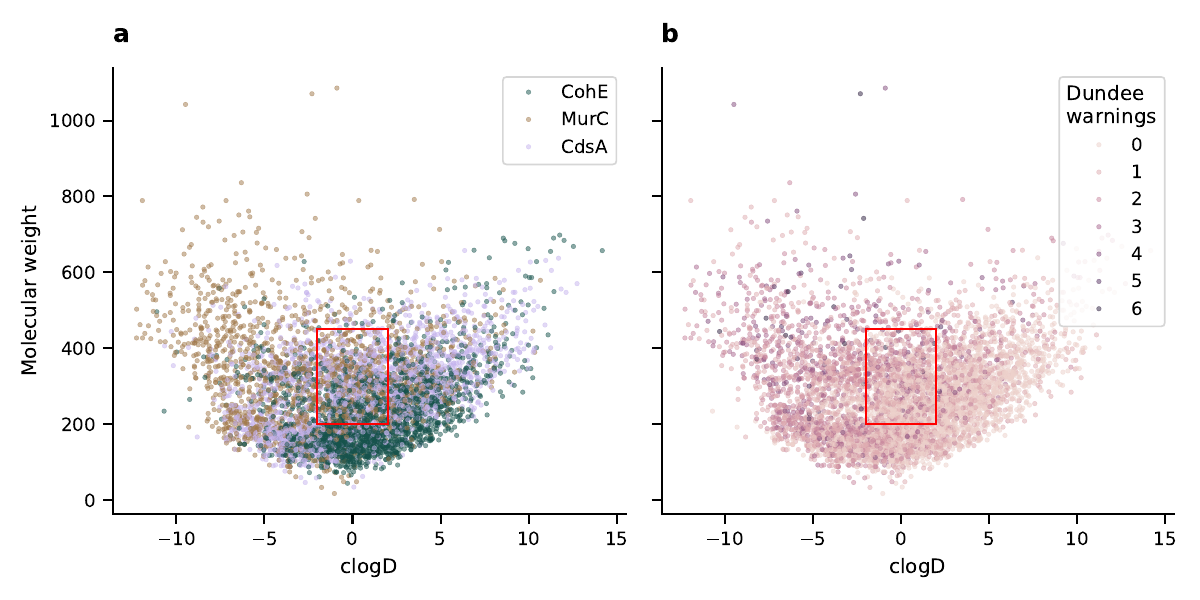}
    \caption{\textbf{Property space in relation to target space and structural alerts.} 
    We investigated the distribution in property space dependant on the target (\textbf{a}) and the number of Dundee alerts (\textbf{b}). 
    Known cytoplasmic antibiotics property space is marked as a \textcolor{red}{box}. 
    We visualized molecular weight and calculated logD at pH 7.4. 
    \num{10000} molecules were sampled. 
    Only valid molecules were considered. }
    \label{fig:scatter_target_dundee}
\end{figure}

\end{document}